\documentclass[aps,prd,10pt,preprintnumbers,nofootinbib,superscriptaddress,showpacs]{revtex4-2}

\usepackage{graphicx}
\usepackage{tikz}
\usepackage{bm}
\usepackage{amsmath,amssymb}
\usepackage[colorlinks=true,linkcolor=blue,citecolor=blue,urlcolor=blue]{hyperref}
\usepackage{multirow}
\usepackage[section] {easy-todo}
\usepackage{centernot}
\usepackage{autobreak}

\newcommand{\tr}{\mathrm{tr}}
\newcommand{\trans}{\mathrm{T}}

\begin{document}

\title{Renormalization Group Running of the Minimal Leptophilic Dark Matter Model toward a UV Completion}
\author{Osamu Seto}
\affiliation{Department of Physics, Hokkaido University, Sapporo 060-0810, Japan}
\author{Tetsuo Shindou}
\affiliation{Division of Liberal-Arts, Kogakuin University, 2665-1 Nakano-machi, Hachioji, Tokyo, 192-0015, Japan}
\author{Takanao Tsuyuki}
\affiliation{Faculty of Business Administration and Information Science, Hokkaido Information University, 59-2 Nishi-Nopporo, Ebetsu, Hokkaido 069-8585, Japan}

\preprint{EPHOU-25-016}
\preprint{KU-PH-038}

\date{\today}
\begin{abstract}
	We study the renormalization group running of the coupling constants in a minimal leptophilic dark matter model in which a Standard Model singlet fermion, 
	acting as the dark matter (DM) candidate, couples exclusively to right-handed charged leptons via a new charged scalar mediator. 
	Reproduction of the observed thermal relic abundance of the DM candidate requires sizable Yukawa couplings,
	and such sizable Yukawa couplings can significantly affect the renormalization group evolution of the model parameters.
	We examine the conditions that the model remains perturbative and the vacuum stability is maintained up to high-energy scales.
    We find that the parameter space is severely constrained to ensure the perturbativity and the vacuum stability up to the Planck scale.
	In particular, the masses of the dark matter and the charged scalar mediator should be smaller than about 350~GeV and can be tested by future collider experiments.
    The lower bound of the dark matter mass that is larger than a few GeV is also obtained by perturbativity.
\end{abstract}

\maketitle
\section{Introduction}
The nature of dark matter (DM) remains one of the more serious problems in particle physics and cosmology.
There is no appropriate candidate for the DM in the Standard Model (SM) of particle physics, and 
we need to extend the SM by introducing a new particle that behaves as the DM.

Many ideas of the DM scenarios have been proposed in the literature.
Among such scenarios, leptophilic models, where the DM interacts exclusively with leptons, 
provide a particularly intriguing class of scenarios~\cite{Krauss:2002px,Baltz:2002we,Fox:2008kb,Arkani-Hamed:2008hhe,Dev:2013hka,Bell:2014tta,Bai:2014osa,Duan:2017pkq,Chang:2014tea,Junius:2019dci,Okawa:2020jea,Kawamura:2020qxo,Horigome:2021qof}.
In some models, the DM interacts with the extra $Z'$ gauge boson as a mediator.
In another subclass of the models, no extra gauge boson is introduced, and the Yukawa couplings give the DM interactions with a scalar mediator.
If the DM is a SM singlet fermion, the cross section of DM-nucleus scattering is significantly suppressed because of the absence of the gauge boson mediation, 
while the DM annihilation cross section via the new charged scalar mediation can be large enough to reproduce the observed relic abundance of DM.
Thus, the model can easily avoid the stringent constraints from the direct detection experiments~\cite{LZ:2024zvo}.

In addition, these setups can be realized in radiative neutrino mass generation models, 
such as the well-known Krauss-Nasri-Trodden model~\cite{Krauss:2002px}. 
In these frameworks, the same new particles responsible for leptophilic DM interactions can participate in loop-induced neutrino mass generation, 
providing a unified picture of dark matter and neutrino physics.

Despite their phenomenological attractiveness, most studies of leptophilic DM remain at the effective model level, without exploring their ultraviolet (UV) completions.
To gain a more fundamental understanding, it is crucial to investigate the renormalization group (RG) behavior of the coupling constants in the Lagrangian.
To reproduce the correct thermal relic abundance of the DM candidate, sizable Yukawa couplings are required in the model, and these Yukawa couplings can significantly affect the RG behavior
of the Lagrangian parameters.
In addition, the scalar self-couplings in the model also make a significant contribution to RG behavior. 
In particular, they can lead to nontrivial RG running and potential issues with vacuum stability or perturbativity at high energies.
These kinds of vacuum stability bound and a triviality bound have been precisely studied in multi-Higgs doublet models~\cite{Flores:1982pr,Kominis:1993zc,Nie:1998yn,Kanemura:1999xf,Ferreira:2009jb}. 

In this work, we investigate a minimal leptophilic DM model with a singlet fermion dark matter particle and a charged scalar mediator
that couples exclusively to right-handed charged leptons. 
Our focus is to analyze the RG evolution of the couplings and determine the theoretical consistency of the model up to high-energy scales. 
If a coupling becomes nonperturbative or leads to vacuum instability at a certain energy scale, it would suggest that there is a cutoff scale 
beyond which the minimal leptophilic DM model as an effective theory description breaks down.
Note that the cutoff scale is not necessarily identical to the Planck scale or the grand unified theory scale. 
If the cutoff scale is lower than them, we can expect that the model is switched to an intermediate UV theory.
The Krauss-Nasri-Trodden model mentioned above is an example of such an intermediate theory.

The paper is organized as follows. The model is introduced in Sec.~\ref{sec:model}, 
where we discuss the thermal relic abundance of the DM candidate and the necessary size of the Yukawa couplings.
In Sec.~\ref{sec:RGEs}, we investigate the RG running of the Yukawa couplings and the scalar quartic couplings,
and discuss the appearance of the Landau pole and vacuum instability.
In Sec .~\ref {sec:conclusion}, we summarize our findings and discuss the implications of our results.

\section{A model with lepthophilic dark matter}\label{sec:model}
We consider a model with a DM candidate which only couples to the right-handed charged leptons.
In the simplest model, a SM singlet fermion $N$ and a charged $\mathrm{SU}(2)_L$ 
singlet scalar $S^{\pm}$ are introduced to the SM particle content.
We also impose a $\mathrm{Z}_2$ symmetry under which the DM candidate $N$ and the scalar $S^{\pm}$ are odd, while all the other SM particles are even.
The Lagrangian is given by 
\begin{align}
	\mathcal{L}= & \
	\mathcal{L}_{\text{SM}}
	-\left(
		y_i\bar{e}_{Ri}N^cS^-
		+\frac{1}{2}(M_N)\bar{N}N^c
		+\text{h.c.}
	\right)
  -\mu_S^2|S^+|^2-\frac{\lambda_S}{4}(S^+S^-)^2-\lambda_{SH}S^+S^-|H|^2\;.
\end{align}
The physical mass of the charged scalar $S^{\pm}$ is given by
\begin{align}
	m_S^2 = \mu_S^2 + \frac{\lambda_{SH}}{2}v^2\;,
\end{align}
where $v\simeq 246$~GeV is the vacuum expectation value.
The physical mass of the singlet fermion $N$ is $m_N=M_N$.

As the singlet fermion $N$ is the DM candidate in this model, $m_N<m_S$ should be satisfied. 
We need to reproduce the correct thermal relic abundance of the DM candidate $N$ as 
$\Omega_N h^2 \simeq 0.1$.
The abundance is approximately determined by the annihilation cross section of $\sigma v$ as
\begin{align}
	\Omega_N h^2 \simeq 0.1\left(\frac{3\times 10^{-26}\,\mathrm{cm}^3/\mathrm{s}}{\sigma v}\right)\;,
\end{align}
so that the annihilation cross section should be $\sigma v \simeq 3\times 10^{-26}\,\mathrm{cm}^3/\mathrm{s}$.
In most parameter region, the annihilation cross section is dominated by the process 
of $NN \to e_i^+e_j^-$, which is mediated by the charged scalar $S^{\pm}$.
For this process, the annihilation cross section is given by
\begin{align}
	\sigma v \simeq \frac{m_N^2(m_N^4+m_S^4)}{8\pi (m_N^2+m_S^2)^4} x_f
	\sum_i\sum_j|y_i^*y_j|^2\;,
	\label{eq:annihilation:NNtoee}
\end{align}
where $x_f\equiv T_f/m_N$ is determined by the freeze-out temperature of $N$, $T_f$ and 
we take $x_f \simeq 1/20$.
In the case of $m_N\simeq m_S$, the coannihilation process with $S^{\pm}$ 
that gives a significant contribution to the annihilation cross section is taken into account.

\section{renormalization Group running}\label{sec:RGEs}
We focus on the RG running of the Yukawa coupling $y_i$ and 
the scalar quartic couplings, $\lambda_H$, $\lambda_S$, and $\lambda_{SH}$,
where $\lambda_H$ denotes the quartic coupling of the SM Higgs doublet $H$.
The full set of the one-loop beta functions for the Lagrangian parameters in the simplest Leptophilic DM model is given in the Appendix.
In the analysis below, we take the input values for the SM parameters as 
$g_S(m_Z)=1.21$, $g(m_Z)=0.648$, $g_Y(m_Z)=0.358$, 
$y_t(m_Z)=0.978$, 
$y_b(m_Z)=0.0165$,
$y_{\tau}(m_Z)=0.0102$, 
$\mu_H^2(m_Z)=-(88.6~\mathrm{GeV})^2$, and $\lambda_H(m_Z)=0.259$
at the scale of $m_Z=91.2~\mathrm{GeV}$.

Equation~\eqref{eq:annihilation:NNtoee} suggests that 
larger Yukawa couplings $y_i$ are necessary to reproduce the correct thermal relic abundance of the DM candidate $N$ for larger $m_N$ and $m_S$.
If the Yukawa coupling $y_i$ is of the order of $1$, 
its contribution to the RG running of $y_i$ itself and $\lambda_S$ can be significant.
The Yukawa couplings $y_i$ contribute to the beta function of $\lambda_S$ with the quartic power.
Thus, the $\lambda_S$ quickly blows up and the Landau pole will appear once $|y_i(\mu)|$ is larger than $\sqrt{4\pi}$.
On the other hand, the scalar quartic couplings contribute to the one-loop beta functions of any couplings at most with the quadratic power, and then the Landau pole will appear just above the scale where $\lambda_i$ reaches $4\pi$.
In this paper, we set a criterion for the breakdown of perturbativity as $|y_i(\mu)|>\sqrt{4\pi}$ or $\lambda_i(\mu)>4\pi$.

We discuss the running behavior of the Yukawa coupling $y_i$.
The large Yukawa coupling tends to lead to the Landau pole at an energy scale lower than the Planck scale. 
The existence of the Landau pole at the energy scale $\mu_L$ suggests 
the cutoff scale of the theory, $\Lambda_{\mathrm{cut}}$, 
is less than $\mu_L$.

Let us at first analytically estimate the maximal value of 
the cutoff scale $\Lambda_{\mathrm{cut}}^{\mathrm{max}}$
suggested by the RG running of the Yukawa coupling $y_i$.
In this paper, we focus on the case that the only tau lepton couples to $N$, i.e., $y_1=y_2=0$ not to induce  a lepton flavour violation process through $N$ mediation.
The renormalization group equation (RGE) for $y=y_3$ is approximately given by 
\begin{equation}
	\frac{dy}{d\log\mu} = \frac{1}{(4\pi)^2}(-3g_Y^{2}y+2y^3)\;,
\end{equation}
where we ignore the charged lepton Yukawa couplings in Eq. \eqref{eq:betay}.
By using the RGE for $U(1)_Y$ gauge coupling $g_Y$ 
[$\mu>m_S$ in Eq. \eqref{eq:betag}], 
\begin{align}
    \frac{dg_Y}{d\log\mu} &= \frac{1}{(4\pi)^2}\frac{43}{6}g_Y^{3},
\end{align}
we obtain
\begin{align}
    \frac{dy}{dg_Y}=\frac{6}{43}\left(-3\frac{y}{g_Y}+2\frac{y^3}{g_Y^3}\right).
\end{align}
This is a homogeneous differential equation, and can be integrated as 
\begin{align}
    y(\mu)&=\frac{g_Y(\mu)}{\sqrt{\left(\frac{g_{Y}^2(m_S)}{y^2(m_S)}-\frac{61}{12}\right)\left(\frac{g_Y(\mu)}{g_{Y}(m_S)}\right)^{\frac{122}{43}}+\frac{61}{12}}}, \label{eq:yNRGE:approx} \\
    g_Y(\mu)&=\frac{g_{Y}(m_S)}{\sqrt{1-\frac{43}{48}g_{Y}^2(m_S)\log(\mu/m_S)}}.
\end{align}
Here, we can approximately evaluate $\Lambda_{\mathrm{cut}}^{\mathrm{max}}$ by the condition of $y(\Lambda_{\mathrm{cut}}^{\mathrm{max}})=\sqrt{4\pi}$.

\begin{figure}[t]
	\includegraphics[width=0.4\textwidth]{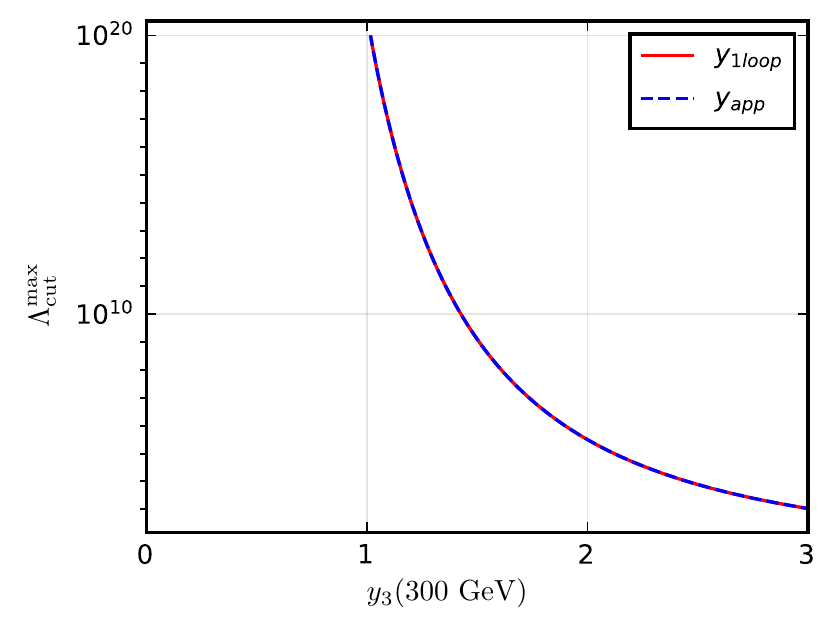}
	\caption{
		The value of $\Lambda_{\mathrm{cut}}^{\mathrm{max}}$ as a function of the 
		input Yukawa coupling $y_3$ at $\mu=300$~GeV.
		The solid (red) curve shows the numerical solution of the one-loop RGE for $y_3$ in the 
		case that $N$ only couples to the tau lepton, while the dashed (blue) curve shows the
		approximate solution given in Eq.~\eqref{eq:yNRGE:approx}.
		The two curves are almost identical.}
	\label{figyNRGE}
\end{figure}

In Fig.~\ref{figyNRGE}, 
the value of $\Lambda_{\mathrm{cut}}^{\mathrm{max}}$ is shown as a function of the input of $y_i$ at the scale $\mu=300$~GeV 
for both the approximate solution given in Eq.~\eqref{eq:yNRGE:approx} and the numerical solution of the one-loop RGE.
In the numerical computation, we consider the case that $N$ only couples to the tau lepton. 
If one wants to consider the model valid up to the Planck scale, 
the Yukawa coupling $y$ should be smaller than about $1$ at $\mu=300$~GeV.

Large $y$ contributes not only to the RGE of $y$ itself but also to the RGE of the scalar quartic couplings.
In particular, there is a negative contribution 
in the beta function of $\lambda_S$ [see Eq. \eqref{eq:betals}] with a large coefficient as 
$-8\mathrm{tr}(y^{\dagger} y y^{\dagger}y)$.
This contribution pulls down $\lambda_S$ in the running up and can easily drive it negative, which leads to terrible vacuum instability.
To compensate for the negative contribution, 
we need to take a large value of $\lambda_S$ or $\lambda_{SH}$. 
However, large $\lambda_S$ and $\lambda_{SH}$ in turn cause 
the blow up of the quartic couplings $\lambda_H$, $\lambda_S$, and $\lambda_{SH}$, and 
lead to the Landau pole, which breaks down the perturbative effective theory description.

\begin{figure}[t]
\begin{tabular}{cc}
	\includegraphics[width=0.32\textwidth]{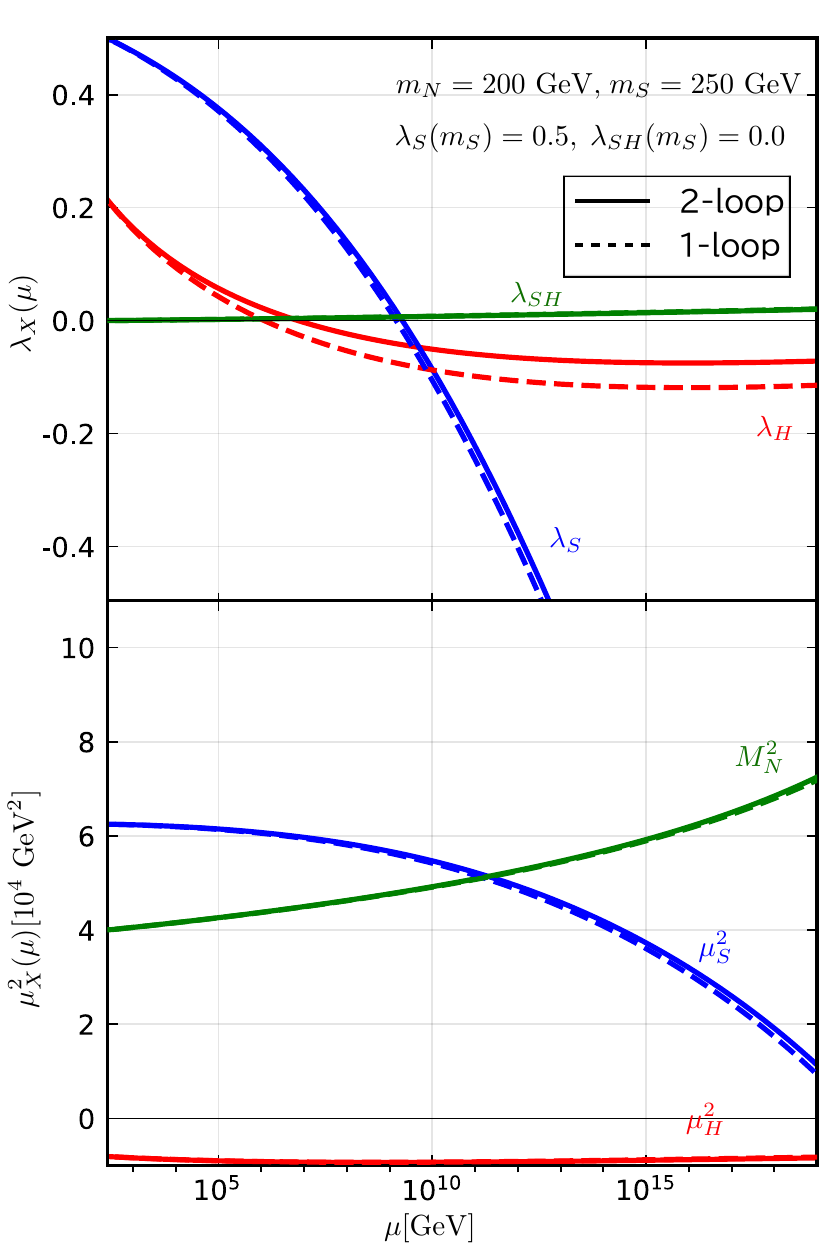}&
	\includegraphics[width=0.32\textwidth]{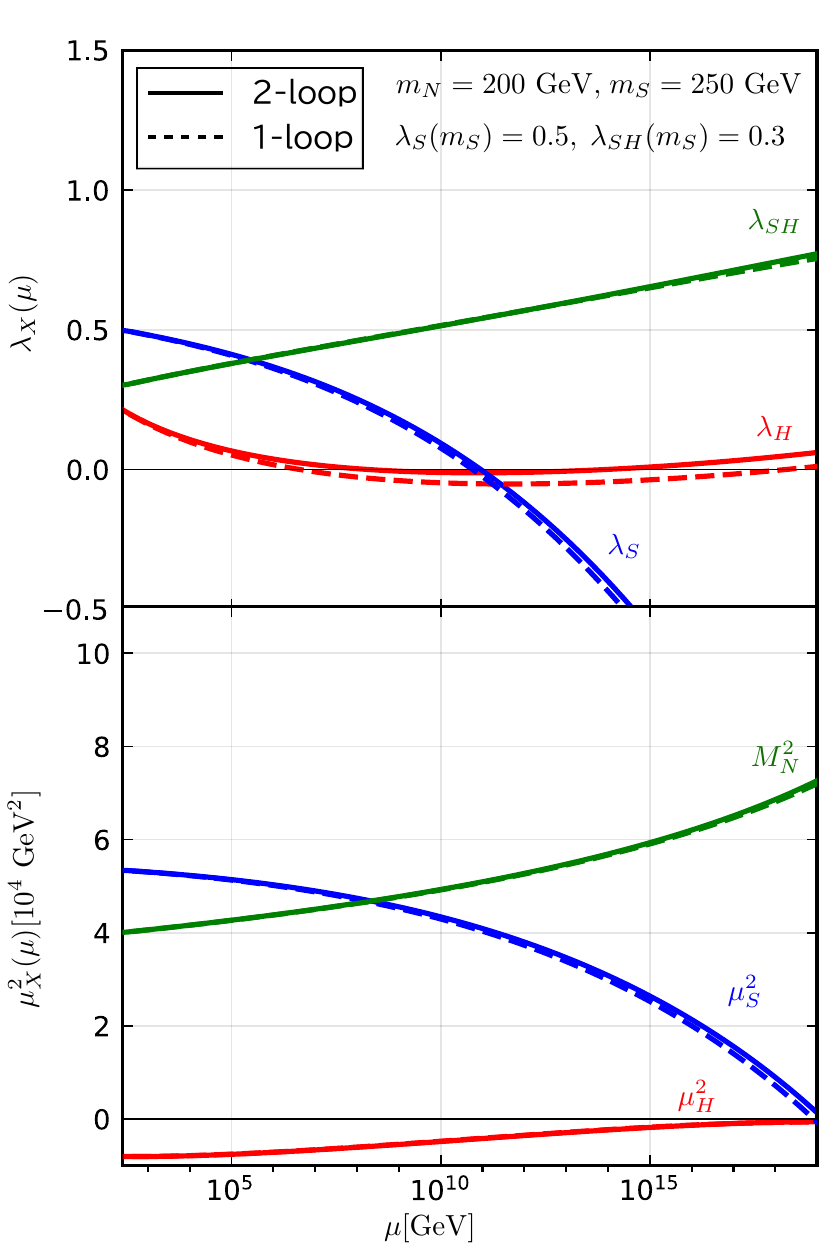}\\
	(a)&(b)\\
	\includegraphics[width=0.32\textwidth]{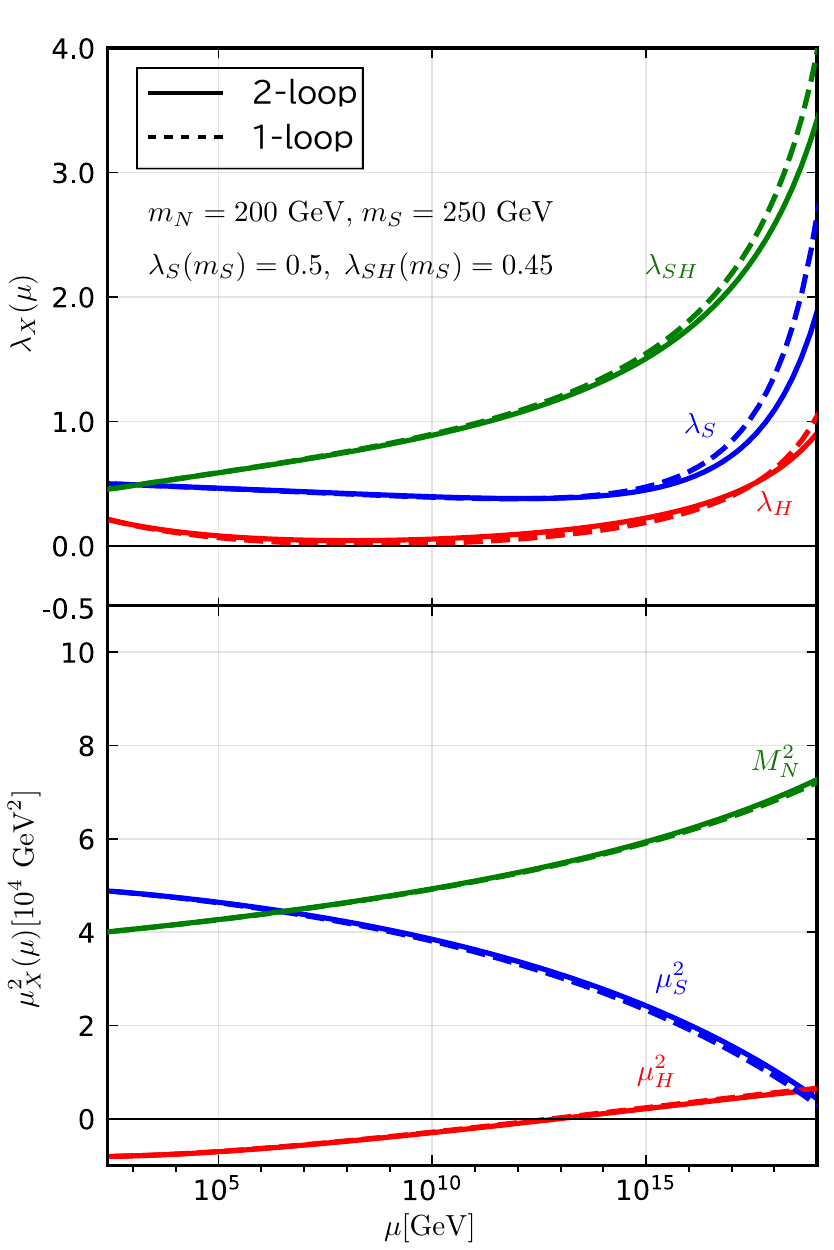}&
	\includegraphics[width=0.32\textwidth]{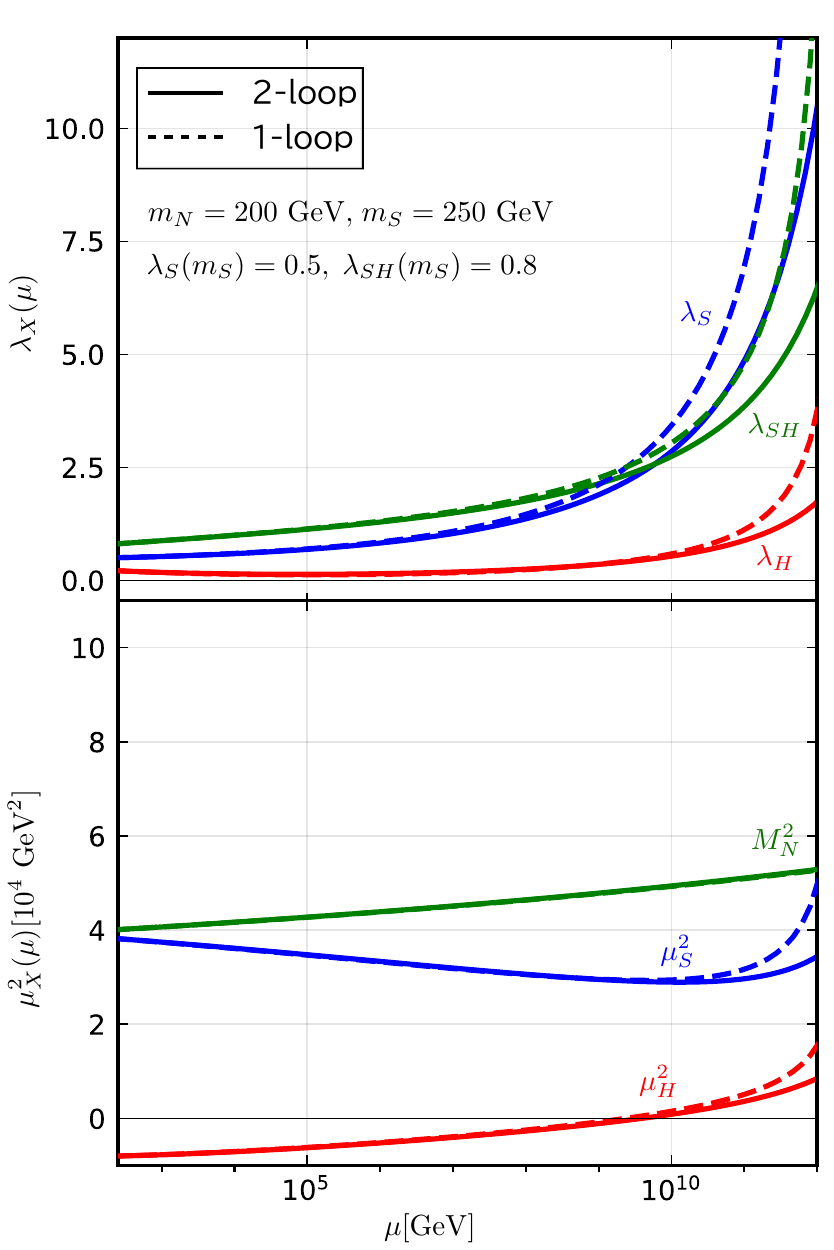}\\
	(c)&(d)\\
\end{tabular}
\caption{
		The RG running of the scalar quartic couplings $\lambda_H$, $\lambda_S$, and $\lambda_{SH}$, 
		and the mass squared parameters $\mu_H^2$ and $\mu_S^2$ 
		in the case that $m_N=200$~GeV, $m_S=250$~GeV, $\lambda_S(m_S)=0.5$.
		The panels (a), (b), (c), and (d) correspond to the initial values of $\lambda_{SH}(m_S)=0.0, 0.3, 0.45, 0.8$, respectively.
		In each panel, the dashed curves show the running with the one-loop beta functions, while the solid curves show the running with the two-loop beta functions.
	}
	\label{fig:RGEs:mN200mS250lamS0.5}
\end{figure}

\begin{figure}[t]
\begin{tabular}{cc}
	\includegraphics[width=0.32\textwidth]{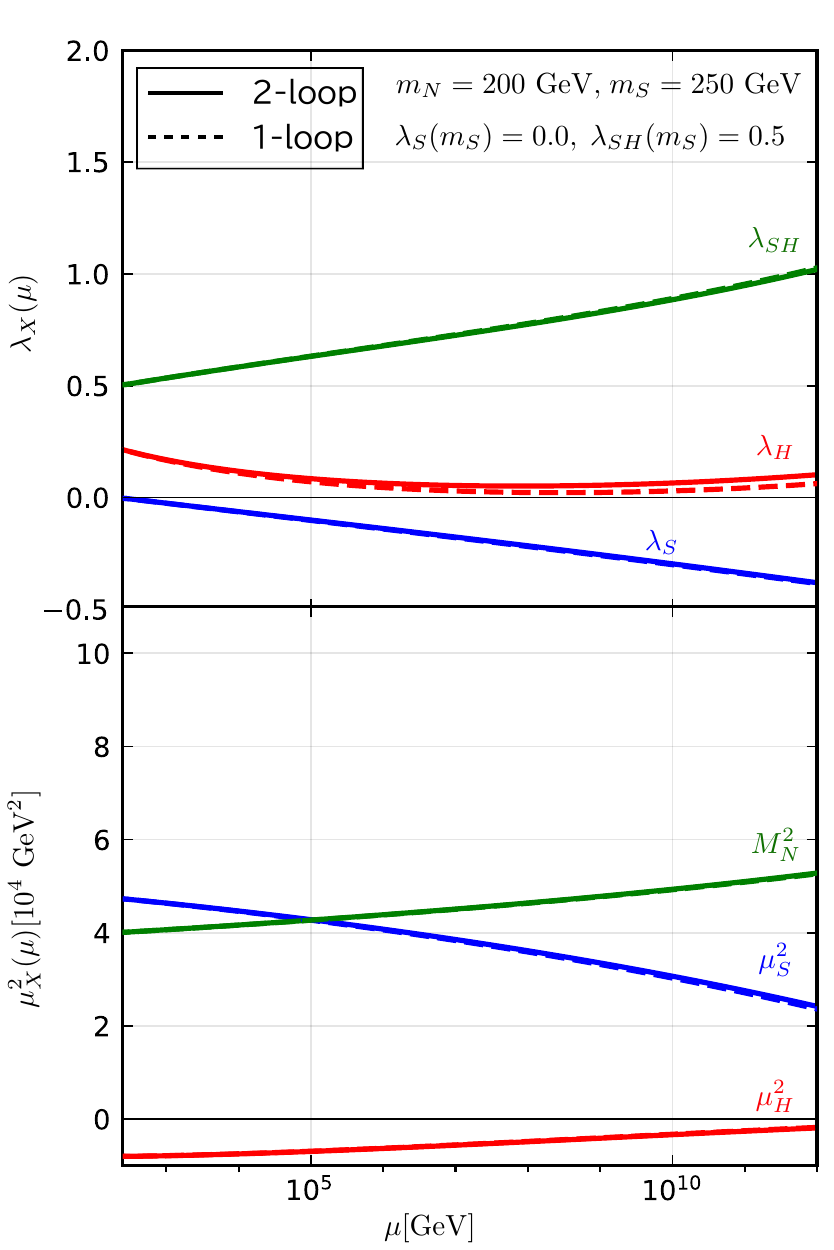}&
	\includegraphics[width=0.32\textwidth]{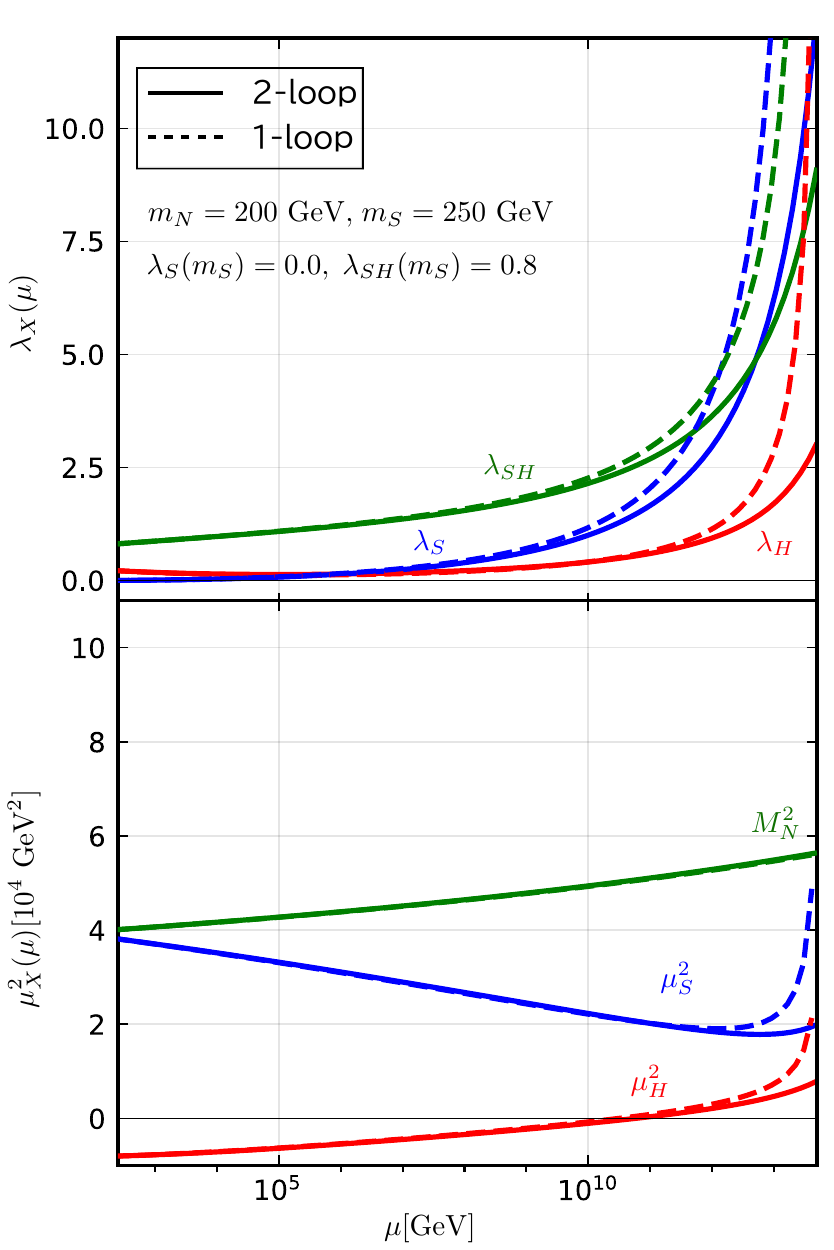}\\
    	(a)&(b)\\
\end{tabular}
\caption{
		The RG running of the scalar quartic couplings $\lambda_H$, $\lambda_S$, and $\lambda_{SH}$, 
		and the mass squared parameters $\mu_H^2$ and $\mu_S^2$ 
		in the case that $m_N=200$~GeV, $m_S=250$~GeV, $\lambda_S(m_S)=0$.
		The panels (a) and (b) correspond to the initial values of $\lambda_{SH}(m_S)=0.5$ and $0.8$, respectively.
		In each panel, the dashed curves show the running with the one-loop beta functions, while the solid curves show the running with the two-loop beta functions.
	}
	\label{fig:RGEs:mN200mS250lamS0.0}
\end{figure}

\begin{figure}[t]
\begin{tabular}{cc}
	\includegraphics[width=0.32\textwidth]{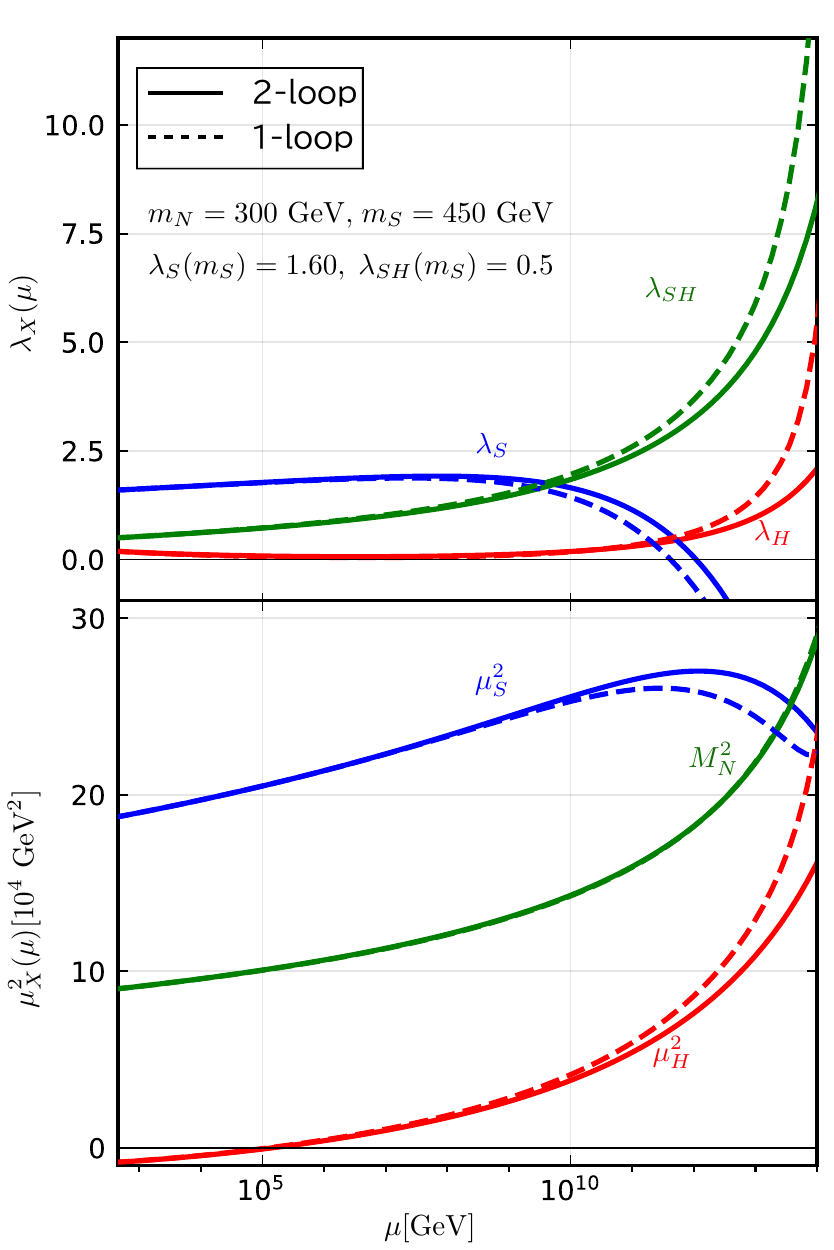}&
	\includegraphics[width=0.32\textwidth]{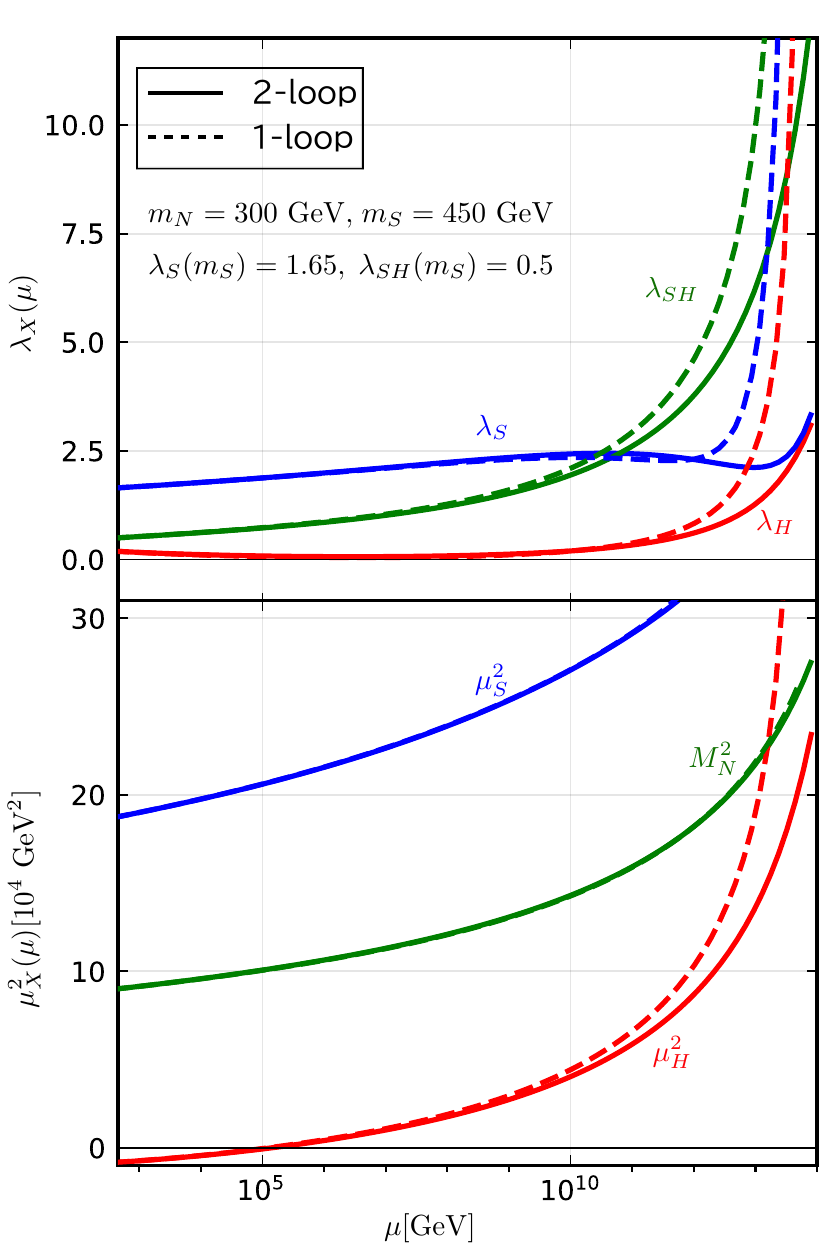}\\
	(a)&(b)\\
\end{tabular}
\caption{
		The RG running of the scalar quartic couplings $\lambda_H$, $\lambda_S$, and $\lambda_{SH}$, 
		and the mass squared parameters $\mu_H^2$ and $\mu_S^2$ 
		in the case that $m_N=300$~GeV, $m_S=450$~GeV, $\lambda_{SH}(m_S)=0.5$.
		The panels (a) and (b) correspond to the initial values of $\lambda_{S}(m_S)=1.60$ and $1.65$, respectively.
		In each panel, the dashed curves show the running with the one-loop beta functions, while the solid curves show the running with the two-loop beta functions.
	}
	\label{fig:RGEs:mN300mS450lamSH0.5}
\end{figure}

In Figs.~\ref{fig:RGEs:mN200mS250lamS0.5}--\ref{fig:RGEs:mN300mS450lamSH0.5},
the RG running of the scalar quartic couplings $\lambda_H$, $\lambda_S$, and $\lambda_{SH}$,
and the mass squared parameters $\mu_H^2$ and $\mu_S^2$ are displayed.
For $m_N=200$~GeV and $m_S=250$~GeV, $y_3\simeq 0.9$ is required to reproduce the correct thermal relic abundance of the DM candidate $N$.
For a rather small value of $\lambda_{SH}(m_S)$, 
the negative contribution from the Yukawa coupling $y_3$ to the RGE of $\lambda_S$ overcomes the positive contribution from the quartic coupling $\lambda_{SH}$,
and the value of $\lambda_S$ becomes negative at a certain energy scale and the vacuum instability occurs, as shown in Figs.~\ref{fig:RGEs:mN200mS250lamS0.5}(a) and \ref{fig:RGEs:mN200mS250lamS0.5}(b).
On the other hand, for a larger value of $\lambda_{SH}(m_S)$, 
the $\lambda_S$ is pulled up by the positive contribution from $\lambda_{SH}$.
If the positive contribution is too large, the Landau pole appears at a lower energy scale than the Planck scale, as shown in Fig.~\ref{fig:RGEs:mN200mS250lamS0.5}(d).
As shown in Fig.~\ref{fig:RGEs:mN200mS250lamS0.0}, 
even in the case of $\lambda_S(m_S)=0.0$, the Landau pole tends to appear, though the vacuum instability can be avoided by taking a large value of $\lambda_{SH}(m_S)$.

The size of Yukawa coupling $y_3$ becomes larger for larger values of $m_N$ and $m_S$ to explain the DM relic abundance.
A case with $m_N=300$~GeV and $m_S=450$~GeV requires $y_3\simeq 1.2$.
We need to take larger $\lambda_S(m_S)$ or larger $\lambda_{SH}(m_S)$ to avoid the vacuum instability,
and the property of the RG running is much more sensitive to the initial value of $\lambda_S(m_S)$ and $\lambda_{SH}(m_S)$ 
for a larger value of Yukawa coupling $y_3$.
Comparing (a) and (b) in 	Fig.~\ref{fig:RGEs:mN300mS450lamSH0.5}, 
we find how the RG running depends on the initial value of $\lambda_S(m_S)$.
It shows that the value of $\lambda_S$ is smaller than a certain value, 
$\lambda_S$ is pulled down to be negative, while for a bit larger value of $\lambda_S(m_S)$,
$\lambda_S$ is pulled up to break down the perturbative treatment.

It is known that the quartic scalar coupling of the SM Higgs doublet $\lambda_H$ becomes negative at a lower energy scale than the Planck scale
in the SM by the negative contribution from the top Yukawa coupling (see Ref.~\cite{Hiller:2024zjp} for example).
We can find that the $\lambda_H$ can be pulled up and avoid the vacuum instability 
in the case with significant contribution from $\lambda_S$ and $\lambda_{SH}$ as a secondary benefit.
It also should be noted that the radiative breaking of the electroweak symmetry is realized 
in some cases, for example, $m_N=200$~GeV, $m_S=250$~GeV, $\lambda_S(m_S)=0.5$, and $\lambda_{SH}(m_S)=0.45$ 
as shown in Fig.~\ref{fig:RGEs:mN200mS250lamS0.5}(c). 


\begin{figure}[t]
\begin{tabular}{ccc}
	1-loop& 2-loop\\
	\includegraphics[width=0.32\textwidth]{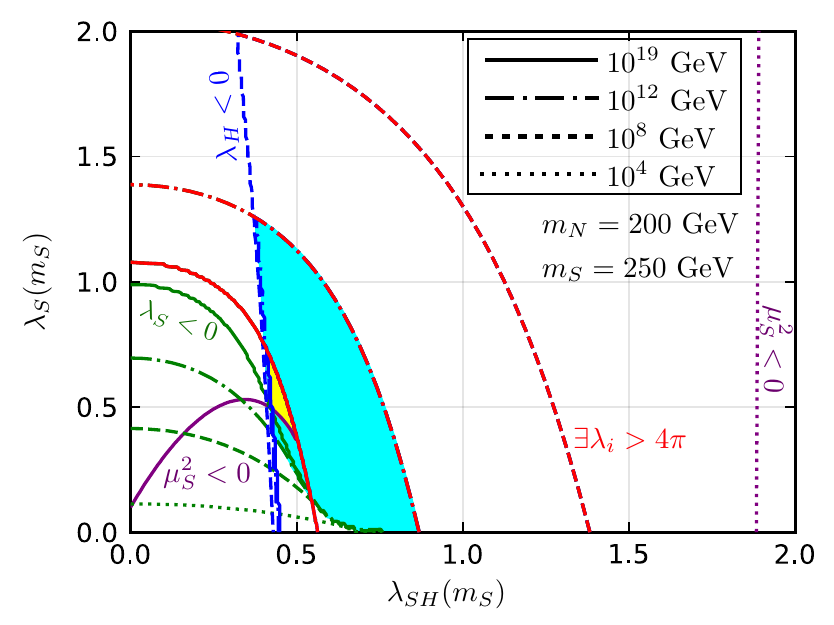}&
	\includegraphics[width=0.32\textwidth]{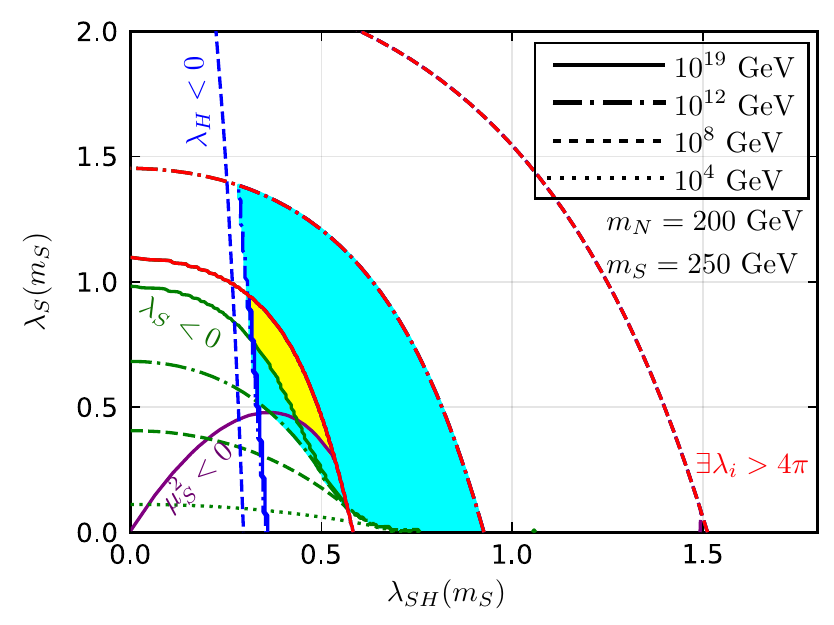}\\
	\multicolumn{2}{c}{(a) $m_N=200$~GeV, $m_S=250$~GeV}\\[5mm]
    1-loop& 2-loop\\
    \includegraphics[width=0.32\textwidth]{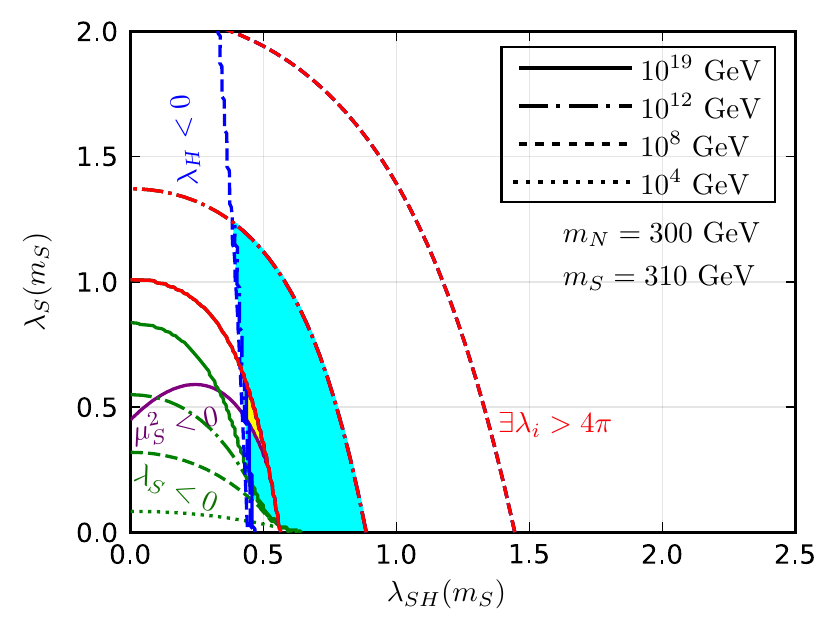}&
	\includegraphics[width=0.32\textwidth]{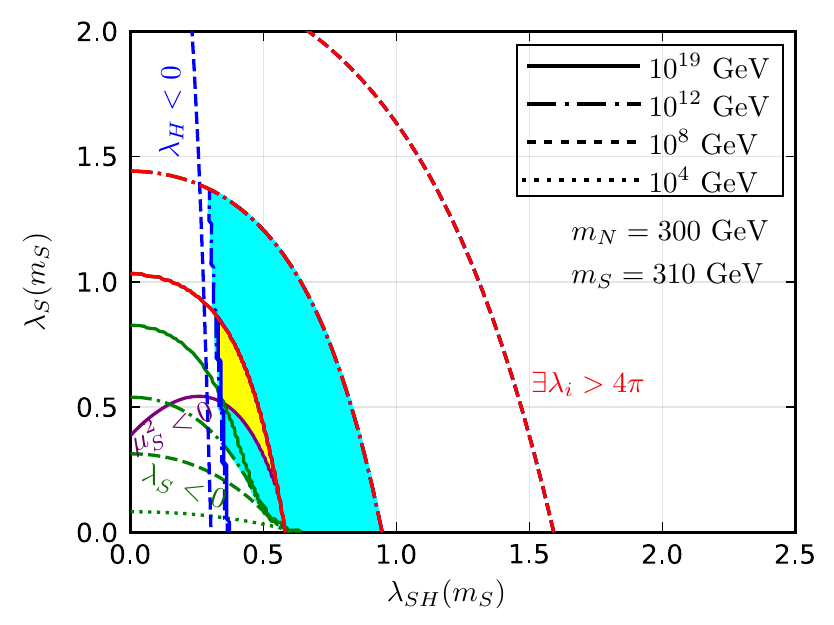}\\
\multicolumn{2}{c}{(b) $m_N=300$~GeV, $m_S=310$~GeV}\\[5mm]
	1-loop& 2-loop\\
	\includegraphics[width=0.32\textwidth]{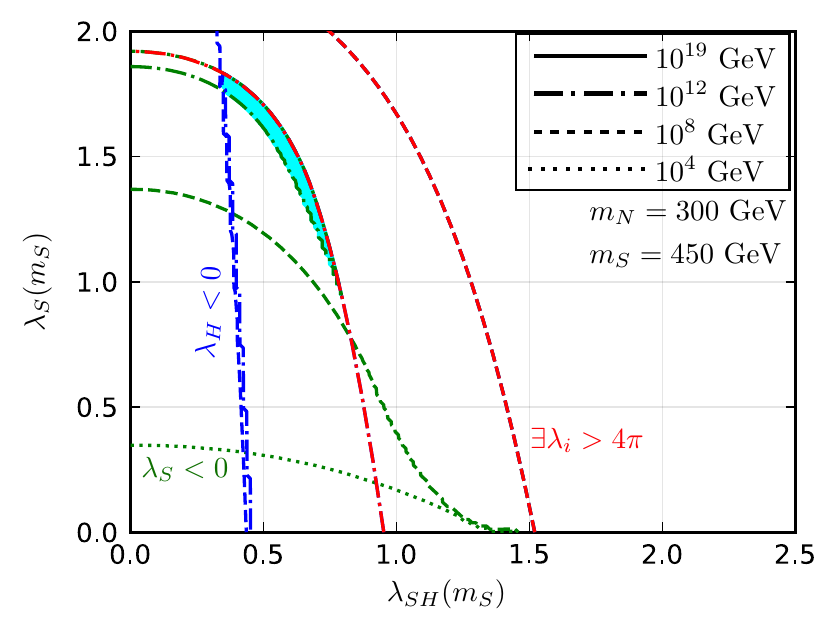}&
	\includegraphics[width=0.32\textwidth]{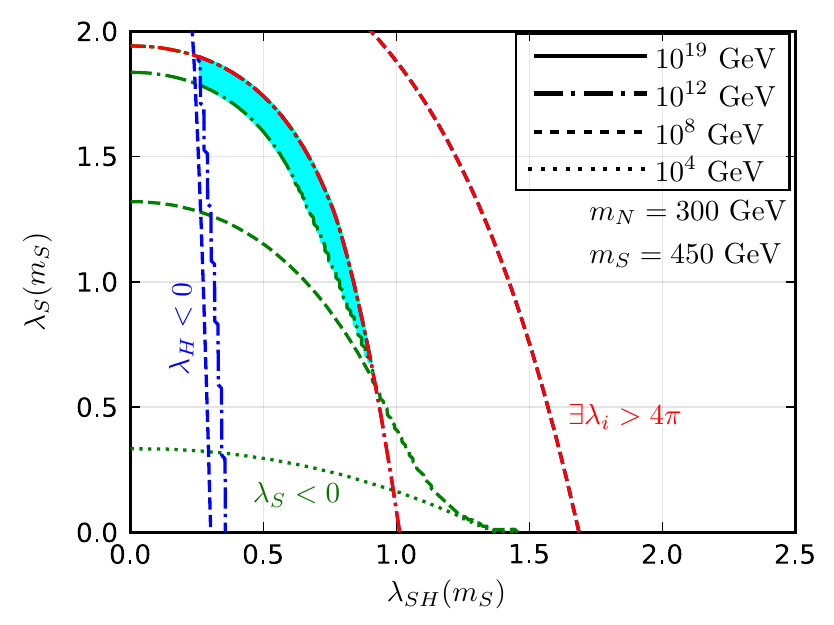}\\
	\multicolumn{2}{c}{(c) $m_N=300$~GeV, $m_S=450$~GeV}\\[5mm]
\end{tabular}
\caption{Contour plot of the energy scale where 
the perturbative effective theory description breaks down.
on the plane of $\lambda_{SH}(m_S)$ and $\lambda_S(m_S)$ for fixed values of $m_N$ and $m_S$.
In each figure, the red curves show the contour of the scale where 
one of the quartic couplings $\lambda_H$, $\lambda_S$, and $\lambda_{SH}$ become nonperturbative (i.e., ${}^{\exists}\lambda_i > 4\pi$).
The blue and green curves show the scale where $\lambda_H<0$ and $\lambda_S<0$, respectively.
The purple curve shows the contour that $\mu_S^2<0$ at $10^{19}~\mathrm{GeV}$ and 
$\mu_S^2<0$ occurs at the lower scale in the bottom-left side range of the curve.
The cyan shaded region shows the range where the effective theory description is valid up to $10^{12}~\mathrm{GeV}$,
and 
the yellow shaded region shows the range where the effective theory description is valid up to $10^{19}~\mathrm{GeV}$.
}
\label{fig:lamSHlamScontour}
\end{figure}

Figure~\ref{fig:lamSHlamScontour} shows the contour plot of the energy scale where 
the perturbative effective theory description breaks down
on the plane of $\lambda_{SH}(m_S)$ and $\lambda_S(m_S)$
for fixed values of $m_N$ and $m_S$.
Either $\lambda_S(m_S)$ or $\lambda_{SH}(m_S)$ becomes larger, 
the Landau pole appears at a lower energy scale, and the perturbative description is broken down.
For small values of $\lambda_{S}(m_S)$ and $\lambda_{SH}(m_S)$,
$\lambda_S$ becomes negative at a significantly low-energy scale.
In cases with 
$(m_N,m_S)=(200~\mathrm{GeV},250~\mathrm{GeV})$ or 
$(m_N, m_S)=(300~\mathrm{GeV},310~\mathrm{GeV})$ which 
corresponds to  $\mu_S^2(m_S)\lesssim |M_N(m_S)|^2$, 
the mass-squared parameter $\mu_S^2$ can be pulled down to negative at a certain 
energy scale 
as seen in Eq.~(\ref{beta_mu_S})
if $\lambda_S(m_S)$ is small.
It is because the positive contribution from the term of $\lambda_{S}\mu_S^2$ to the RGE of $\mu_S^2$ 
is not large enough to compensate the negative contribution from the Yukawa coupling $\mathrm{tr}(y^{\dagger} y)|M_N|^2$ and $\lambda_{SH}\mu_H^2$.

\begin{figure}[t]
	\begin{center}
		\begin{tabular}{cc}
		1-loop&2-loop\\
		\includegraphics[width=0.5\textwidth]{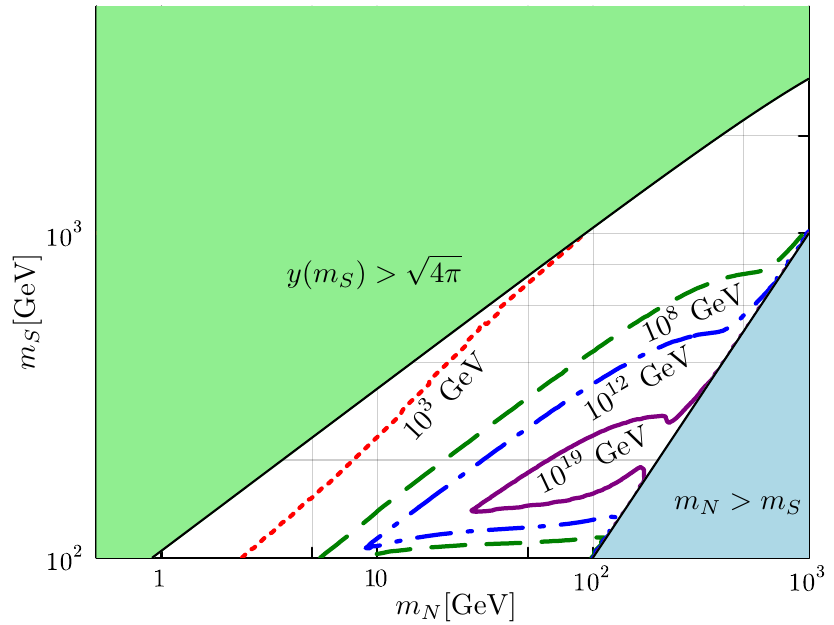}&
		\includegraphics[width=0.5\textwidth]{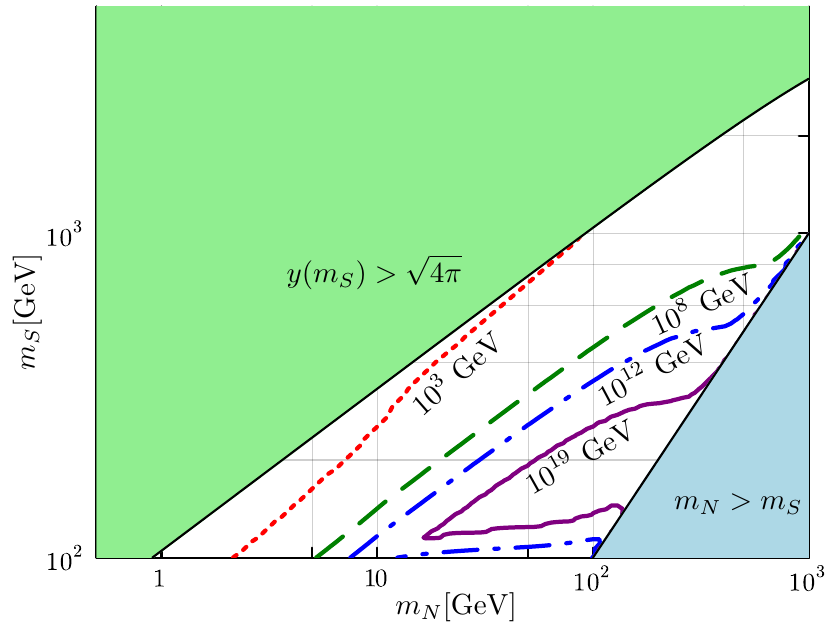}
		\end{tabular}
	\end{center}
	\caption{The contour plot of the maximal value of the cutoff scale $\Lambda_{\mathrm{cut}}^{\mathrm{max}}$
	on the plane of $m_N$ and $m_S$.
	For each point, the size of the Yukawa coupling $y_i$ is tuned to reproduce the correct thermal relic abundance of the DM candidate $N$.
	The right-bottom (light-blue) shaded region is excluded by the condition of $m_N<m_S$ which is required for $N$ being the DM candidate.
	The left-top (light-green) shaded region shows that $y_i>\sqrt{4\pi}$ is required already at $\mu=m_S$.
	The left panel shows the result with the one-loop beta functions, while the right panel shows the result with the two-loop beta functions.
	}
	\label{fig:mNmS}
\end{figure}

In Fig.~\ref{fig:mNmS}, the maximal value of the cutoff scale $\Lambda_{\mathrm{cut}}^{\mathrm{max}}$ is shown as a function of $m_N$ and $m_S$.
The cutoff scale $\Lambda_{\mathrm{cut}}^{\mathrm{max}}$ is estimated by taking into account 
the perturbativity of the $\lambda_H$, $\lambda_S$, $\lambda_{SH}$, and $y$,
and the positivity of $\lambda_H$, $\lambda_S$, $\lambda_{SH}$, and $\mu_S^2$.
In the region of $m_S>150~\mathrm{GeV}$, 
for larger values of $m_S$ and smaller values of $m_N$, 
the maximal value of the cutoff scale $\Lambda_{\mathrm{cut}}^{\mathrm{max}}$ becomes smaller.
When $m_N$ is smaller than $\mathcal{O}(1~\mathrm{GeV})$, 
the nonperturbative value of $y_i>\sqrt{4\pi}$ is required already at $\mu=m_S$ to reproduce the correct thermal relic abundance of the DM candidate $N$, 
and the perturbative description is never valid.
Thus, a sub-GeV thermal DM cannot be realized in the minimal model.
In the case of $m_N\simeq m_S$, which corresponds to the coannihilation region,
the necessary value of the Yukawa coupling is smaller, and the maximal value of the cutoff scale $\Lambda_{\mathrm{cut}}^{\mathrm{max}}$ is not more strongly suppressed than the nondegenerate region.
If one wants to consider the case that this simple model is valid up to the Planck scale,
the DM mass $m_N$ and the mediator mass $m_S$ should be in the range of 
$20~\mathrm{GeV}\lesssim m_N\lesssim 350~\mathrm{GeV}$ and
$150~\mathrm{GeV}\lesssim m_S\lesssim 350~\mathrm{GeV}$.


The stability of the vacuum is a more delicate issue. 
We here test not the stability of the vacuum but only the positivity of the scalar quartic couplings and the mass squared parameter of the charged scalar.
Essentially, a potential analysis including the estimation of the lifetime of our metastable vacuum is necessary to obtain firm constraints. 
Although a precise analysis of the potential would change constraints on parameters, we expect that our conclusion is essentially unchanged. 

\section{Conclusion}\label{sec:conclusion}
In this work, we have studied a minimal leptophilic dark matter model in which a SM-singlet fermion $N$, acting as the DM candidate, 
interacts solely with the right-handed charged leptons via a charged scalar singlet $S^\pm$. 
By imposing a discrete $Z_2$ symmetry, we ensured the stability of the DM particle and analyzed its thermal relic abundance. 
Since the dominant annihilation channel $NN \to e_i^+ e_j^-$ mediated by $S^\pm$
is determined by the Yukawa couplings $y_i$, 
appropriate sizes of the Yukawa couplings are determined for the DM mass $m_N$ and the charged scalar mass $m_S$ to reproduce the observed DM relic density.

We then investigated the RG running of the Yukawa couplings,  the scalar quartic couplings, and mass squared parameters. 
Our analysis reveals that large values of the Yukawa coupling $y_i$ 
which are required for a large DM mass, 
tend to drive the theory to a Landau pole below the Planck scale. 

Moreover, we showed that large Yukawa couplings can lead 
to vacuum instability through their negative contributions 
to the beta function of the quartic coupling $\lambda_S$. 
While increasing $\lambda_S$ or $\lambda_{SH}$ can stabilize the potential,
such choices can make the scalar sector hit a Landau pole, 
limiting the perturbative validity of the model.
The vacuum instability also occurs due to the negative mass-squared parameter $\mu_S^2$, which happens only in a limited parameter region.
The appearance of the Landau pole and the vacuum instability set an upper limit on the cutoff scale of the model.

At the cutoff scale, the model will be switched to a UV theory.
In many cases, new particles are introduced to the model above the cutoff scale.
For example, the leptophilic DM model can be realized
as a low-energy effective theory of the Kraus-Nasri-Trodden model~\cite{Krauss:2002px}, 
where an additional scalar particle and additional singlet fermions appear 
above a certain energy scale.
Such additional particles of course affect the RG running of the Yukawa couplings and the scalar quartic couplings, and the constraints on the 
parameter regions may change accordingly.
However, additional scalar particles give only a positive
contribution to the beta function of $\lambda_S$ and $\lambda_{SH}$.
We naively expect that the Landau pole appears at a lower energy scale than the
simple leptophilic DM model.

Our results highlight a critical tension between the requirements 
of relic abundance, vacuum stability, and perturbativity. 
We have shown how significant the renormalization group running of the scalar potential is 
in constraining viable parameter spaces in dark matter models 
with large Yukawa couplings. 
These findings provide a theoretical 
guideline for building a more fundamental picture of 
the leptophilic DM model.
Our results naively suggested that the DM mass $m_N$ and the mediator mass $m_S$ should be smaller than about $350$~GeV, which can be tested by 
future lepton collider experiments, such as the high-energy option 
of the International Linear Collider~\cite{ILC:2013jhg} and the Compact Linear Collider~\cite{CLICdp:2018cto}.

\begin{acknowledgments}
	This work was supported in part by JSPS KAKENHI Grants No.~23K03402 (O.S.) and No.~20H00160 (T.S.).
\end{acknowledgments}

\appendix
\section{RGEs in Leptophilic Dark Matter Model}\label{app:RGEs:LeptophilicDM}.

The one-loop beta functions 
of the Lagrangian parameters in the simplest leptophilic DM model are given by
\begin{align}
&\beta(g_Y) = \frac{1}{3}\left(
			\frac{41}{2}
			+\theta(S^{\pm})
			\right) g_Y^{3}
			\;,\quad
			\beta(g) =
			- \frac{19}{6} g^{3}\;,\quad
			\beta(g_S) =
			-7 g_S^{3}\;, 
			\displaybreak[0]
			\label{eq:betag}\\
&\beta(Y_u) = 
			-
			\left(
			\frac{17}{12} g_Y^{2}
			-  \frac{9}{4} g^{2}
			- 8 g_S^{2}
			\right) Y_u
			+ \frac{3}{2} Y_u
			\left(
			Y_u^{\dagger} Y_u		-  Y_d^{\dagger} Y_d
			\right)
			+ \left(
			3 \tr\left(Y_u^{\dagger} Y_u \right)
			+ 3 \tr\left(Y_d^{\dagger} Y_d \right)
			+ \tr\left(Y_e^{\dagger} Y_e \right)
			\right)Y_u
			\;,                    
			\displaybreak[0]
			\\
&\beta(Y_d) =
			-  \frac{5}{12} g_Y^{2} Y_d
			-  \frac{9}{4} g^{2} Y_d
			- 8 g_S^{2} Y_d
			-  \frac{3}{2} Y_d Y_u^{\dagger} Y_u
			+ \frac{3}{2} Y_d Y_d^{\dagger} Y_d
			+ \left(
			3 \tr\left(Y_u^{\dagger} Y_u \right)
			+ 3 \tr\left(Y_d^{\dagger} Y_d \right)
			+ \tr\left(Y_e^{\dagger} Y_e \right)
			\right)Y_d
			\;,
			\displaybreak[0]
			\\
&\beta(Y_e) = 
			-  \frac{15}{4} g_Y^{2} Y_e
			-  \frac{9}{4} g^{2} Y_e
			+ \frac{3}{2} Y_e Y_e^{\dagger} Y_e
			+ \frac{1}{2} y y^{\dagger} Y_e
			+ \left(
			3 \tr\left(Y_u^{\dagger} Y_u \right)
			+ 3 \tr\left(Y_d^{\dagger} Y_d \right)
			+ \tr\left(Y_e^{\dagger} Y_e \right)
			\right) Y_e
			\;,
			\displaybreak[0]
			\\
&\beta(y) = 
			- 3 g_Y^{2} y
			+ Y_e Y_e^{\dagger} y
			+ y y^{\dagger} y
			+ \tr\left(y^{\dagger} y \right) y\;, \label{eq:betay}
			\displaybreak[0]
			\\
&\beta(M_N) =
				\frac{1}{2} M_N y^{\dagger} y
				+ \frac{1}{2} y^{\trans} y^{*} M_N\;,				
			\displaybreak[0]
			\\
&\beta(\lambda_H) =
				12 \lambda_H^{2}
				+ 2 \lambda_{SH}^{2}
				- 3 g_Y^{2} \lambda_H
				- 9 g^{2} \lambda_H
				+ \frac{3}{4} g_Y^{4}
				+ \frac{3}{2} g_Y^{2} g^{2}
				+ \frac{9}{4} g^{4}
				+ 12 \lambda_H \tr\left(Y_u^{\dagger} Y_u \right)
				+ 12 \lambda_H \tr\left(Y_d^{\dagger} Y_d \right)
				\nonumber\\
&\phantom{\beta(\lambda_H) = }
				+ 4 \lambda_H \tr\left(Y_e^{\dagger} Y_e \right)
				- 12 \tr\left(Y_u^{\dagger} Y_u Y_u^{\dagger} Y_u \right)
				- 12 \tr\left(Y_d^{\dagger} Y_d Y_d^{\dagger} Y_d \right)
				- 4 \tr\left(Y_e^{\dagger} Y_e Y_e^{\dagger} Y_e \right)\;,
			\displaybreak[0]
				\\
&\beta(\lambda_S) =
				5 \lambda_S^{2}
				+ 8 \lambda_{SH}^{2}
				- 12 g_Y^{2} \lambda_S
				+ 24 g_Y^{4}
				+ 4 \lambda_S \tr\left(y^{\dagger} y \right)
				- 8 \tr\left(y^{\dagger} y y^{\dagger} y \right)\;, \label{eq:betals}
			\displaybreak[0]
				\\
&\beta(\lambda_{SH}) =
				6 \lambda_H \lambda_{SH}
				+ 2 \lambda_{SH} \lambda_S
				+ 4 \lambda_{SH}^{2}
				-  \frac{15}{2} g_Y^{2} \lambda_{SH}
				-  \frac{9}{2} g^{2} \lambda_{SH}
				+ 3 g_Y^{4}
			\displaybreak[0]
				\\
&\phantom{\beta(\lambda_{SH}) = }
				+ 6 \lambda_{SH} \tr\left(Y_u^{\dagger} Y_u \right)
				+ 6 \lambda_{SH} \tr\left(Y_d^{\dagger} Y_d \right)
				+ 2 \lambda_{SH} \tr\left(Y_e^{\dagger} Y_e \right)
				+ 2 \lambda_{SH} \tr\left(y^{\dagger} y \right)
				- 4 \tr\left(Y_e^{\dagger} y y^{\dagger} Y_e \right)\;,
			\displaybreak[0]
				\\
&\beta(\mu_H^2) =
				-  \frac{3}{2} g_Y^{2} \mu_H^2
				-  \frac{9}{2} g^{2} \mu_H^2
				+ 6 \lambda_H \mu_H^2
				+ 2 \lambda_{SH} \mu_S^2
				+ 6 \mu_H^2 \tr\left(Y_u^{\dagger} Y_u \right)
				+ 6 \mu_H^2 \tr\left(Y_d^{\dagger} Y_d \right)
				+ 2 \mu_H^2 \tr\left(Y_e^{\dagger} Y_e \right)\;,
			\displaybreak[0]
				\\
&\beta(\mu_S^2) =
				- 6 g_Y^{2} \mu_S^2
				+ 4 \lambda_{SH} \mu_H^2
				+ 2 \lambda_S \mu_S^2
				+ 2 \mu_S^2 \tr\left(y^{\dagger} y \right)
				- 4 \tr\left(y^{\dagger} y\right)|M_N|^2\;,
                \label{beta_mu_S}
\end{align}
where 
$\dfrac{df}{d\log\mu}=\dfrac{\beta(f)}{(4\pi)^2}$ and 
$\theta(\Phi)=\theta(\mu - M_{\Phi})$.
The two-loop beta function can be provided by applying the formulas given in Ref.~\cite{Luo:2002ti}, 
which are implemented in several public codes.
In this paper, we use the \texttt{PyR@TE3}~\cite{Sartore:2020gou} to obtain the two-loop beta functions.

\bibliography{SST.bib}
\bibliographystyle{utphys}
\end{document}